# Quantum Efficiency of the B-centre in hexagonal boron nitride

Karin Yamamura, Nathan Coste*, Helen Zhi Jie Zeng, Milos Toth, Mehran Kianinia and Igor Aharonovich*

[1] School of Mathematical and Physical Sciences, University of Technology Sydney, Ultimo, New South Wales 2007, Australia
[2] ARC Centre of Excellence for Transformative Meta-Optical Systems, University of Technology Sydney, Ultimo, New South Wales 2007, Australia

* To whom correspondence should be addressed: *nathan.coste@uts.edu.au; igor.aharonovich@uts.edu.au

B-centres in hexagonal boron nitride (hBN) are gaining significant research interest for quantum photonics applications due to precise emitter positioning and highly reproducible emission wavelengths. Here, we leverage the layered nature of hBN to directly measure the quantum efficiency (QE) of single B-centres. The defects were engineered in a 35 nm flake of hBN using electron beam irradiation, and the local dielectric environment was altered by transferring a 250 nm hBN flake on top of the one containing the emitters. By analysing the resulting change in measured lifetimes, we determined the QE of B-centres in the thin flake of hBN, as well as after the transfer. Our results indicate that B-centres located in thin flakes can exhibit QEs higher than 40%. Near-unity QEs are achievable under reasonable Purcell enhancement for emitters embedded in thick flakes of hBN, highlighting their promise for quantum photonics applications.

**Keywords:** hexagonal boron nitride, single photon emitter, quantum efficiency, B-centre

Solid-state single photon emitters (SPEs) have garnered much attention due to their potential in the development of quantum computation, quantum networks and quantum communications[1,2]. SPEs hosted by the two dimensional material hexagonal boron nitride (hBN) are promising due to their high brightness, extreme stability and high Debye-Waller factors[3]. At cryogenic temperatures, their emission into the Zero Phonon Line (ZPL) can reach Fourier-limited linewidths[4,5]. Recently, a specific type of single emitter in hBN, known as the B-centre, has been identified. It emits at a highly reproducible wavelength of 436 nm, and has a reduced susceptibility to spectral diffusion due to the absence of a first-order Stark effect[6,7].

B-centres can be generated on-demand through electron beam irradiation of hBN, making them exceptionally appealing for fundamental research, as well as nanophotonic integration aimed at exploring and exploiting light-matter interactions[4,8,9]. Recent studies include demonstrations of photon indistinguishability, measurements of in-plane emission polarization axes clustered along three crystallographic directions[10], and incorporation of B-centres in nanophotonic devices[9,11-13].

An essential metric that has not yet been established for B-centres is the internal quantum efficiency (QE), which quantifies the fundamental effectiveness of radiative decay from the excited state. QE is a critical factor in assessing the suitability of any light source for practical applications in quantum technologies[14]. However, determining the QE of solid-state sources is often challenging due to their non-uniform surrounding electromagnetic environments, necessitating non-trivial measurement.

In this paper, we directly measure the QE of B-centres by determining the transition dipole moments and radiative emission characteristics in a controlled local dielectric environment. We characterise a set of single emitters engineered in a thin (35 nm) flake of hBN. We then modify the local dielectric environment by transferring an additional, thick (250 nm) flake of hBN on top of the thin flake. The resulting variation in the radiative decay rates of B-centres is measured and used to deduce the QE before and after the transfer.

Generally, the QE of a quantum emitter is defined as the ratio of the radiative decay rate to the total emission decay rate[15-17]:

$$QE = \frac{k_{rad}}{k_{rad} + k_{non-rad}} \qquad (1)$$

Here, $k_{rad}$ and $k_{non-rad}$ are the radiative and non-radiative decay rates, respectively. Photoluminescence (PL) lifetime ($T_1$) measurements under pulsed excitation provide a direct measure of the sum $k_{rad} + k_{non-rad}$. However, determining the QE requires calculating the relative contributions of these terms. The method for determining the QE is based on the fact that the optical environment around the emitter strongly affects the spontaneous emission rate $k_{rad}$ via the local density of states (LDOS), while the non-radiative decay rate $k_{non-rad}$ remains unaffected by LDOS variations[18-21]. In particular, $k_{rad}$ depends on the distance d to the surface of the host medium, the emission dipole orientation, and refractive indices at the interface when d is shorter than emission wavelength ($\lambda$)[22, 23]. Therefore, the local dielectric environment of an emitter can be modified in a controlled way by modifying the distance to the interface within a medium of known refractive index. This induces a change in the observed decay rate and the variation in the measured lifetime can be used to deduce the QE[18, 24] Thus, the QE of an emitter can be calculated using[15]:

$$QE = \frac{(1-\beta)}{(1-\alpha)} \qquad (2)$$

Here, $\alpha = \frac{k_{rad}}{k_{rad}\infty}$ is the simulated ratio of the radiative decay rate of an emitter near the surface and far below the surface. $\beta = \frac{k_{rad} + k_{non-rad}}{k_{rad}\infty + k_{non-rad}}$ is the measured change in total decay rates. The rates $k_{rad}$ and $k_{rad}\infty$ denote the radiative decay rates for the emitter near the surface and far below the surface, respectively.

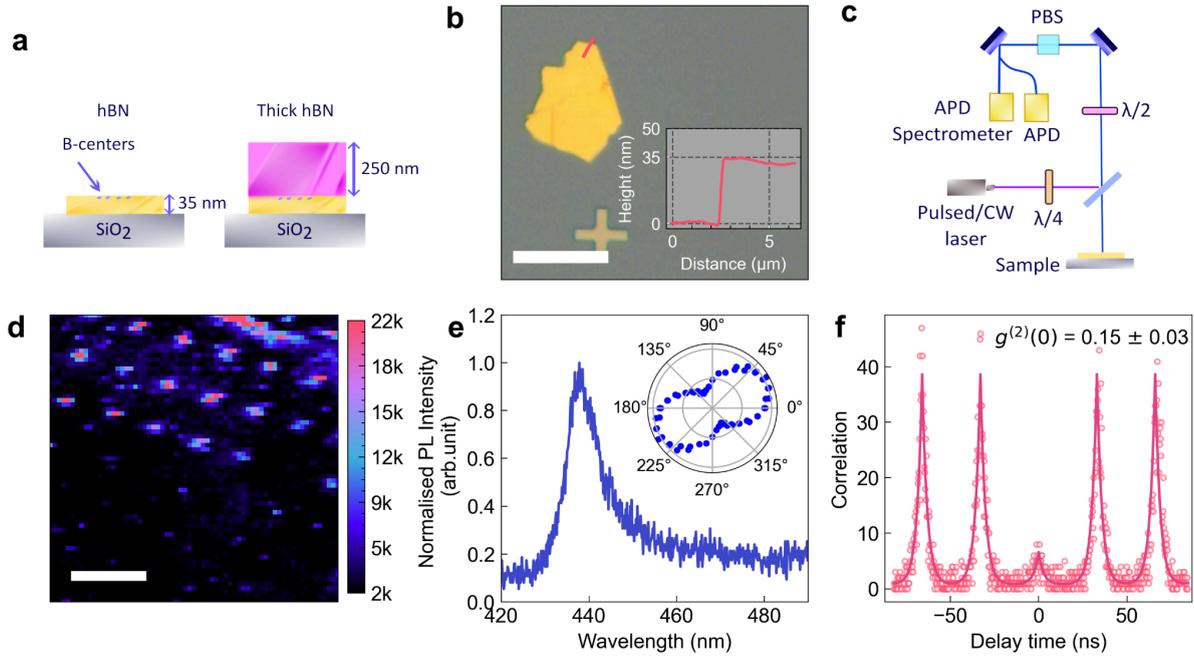

*Figure 1. (a) Schematic of the sample and experimental setup. The B-centres were engineered by electron beam irradiation of a 35 nm flake of hBN. Subsequently, a thick top hBN flake was transferred on the top, functioning as a dielectric medium that modifies the radiative decay rates of the emitters. (b) Optical image of hBN flake and height profile (inset) along the pink line. The scale bar corresponds to 50 µm. (c) Schematics of the experimental setup used for the optical measurements. (d) Confocal PL map showing an array of emissions generated by electron beam irradiation. The scale bar corresponds to 5 µm. (e) PL spectrum from a single B-centre with a ZPL at 436 nm. The inset is a the emission polarization of the B-centre. (f) Second-order photon correlation measurement under pulsed laser excitation of the same emitter, confirming it is a single photon emitter.*

The experimental workflow is illustrated in Figure 1(a). HBN flakes are first exfoliated onto a SiO2 substrate, and a thin ($\ll \lambda$) flake is irradiated by an electron beam to generate an array of B-centres[8, 25]. The SPEs are characterised and the dielectric environment is modified as a thick layer of hBN is transferred on top of the thin flake. The same emitters are then re-characterised in the modified dielectric environment. This technique is advantageous particularly with layered materials, where the top layer (second hBN flake) is having same refractive index as the bottom layer. Figure 4(b) shows an Atomic Force Microscopy (AFM) profile of the thin, 35 nm flake of hBN. All optical characterisation was carried out at room temperature using the confocal PL setup sketched in Figure 1(c). Excitation was performed using a 402 nm pulsed laser with a power of 50 µW and a repetition rate of 40 MHz, and a 405 nm continuous wave laser with a power of 200 µW. A confocal PL map of the B-centre array is shown in Fig. 1(d). A typical PL spectrum showing the B-centre ZPL at 436 nm is shown in Fig. 1(e). The emission is linearly polarized (Fig. 1(e) inset), consistent with an in-plane dipole moment, with a measured visibility, $V = \frac{(p_{max} - p_{min})}{(p_{max} + p_{min})}$, of 0.542. This value is limited primarily by background emission, with previous studies of emission polarization in B-centres reporting polarization visibilities greater than 0.6 at cryogenic temperature. The anti-bunched nature of the single-photon emission is confirmed by the second-order autocorrelation function shown in Fig. 1(f), where $g^{(2)}(0) = 0.15$.

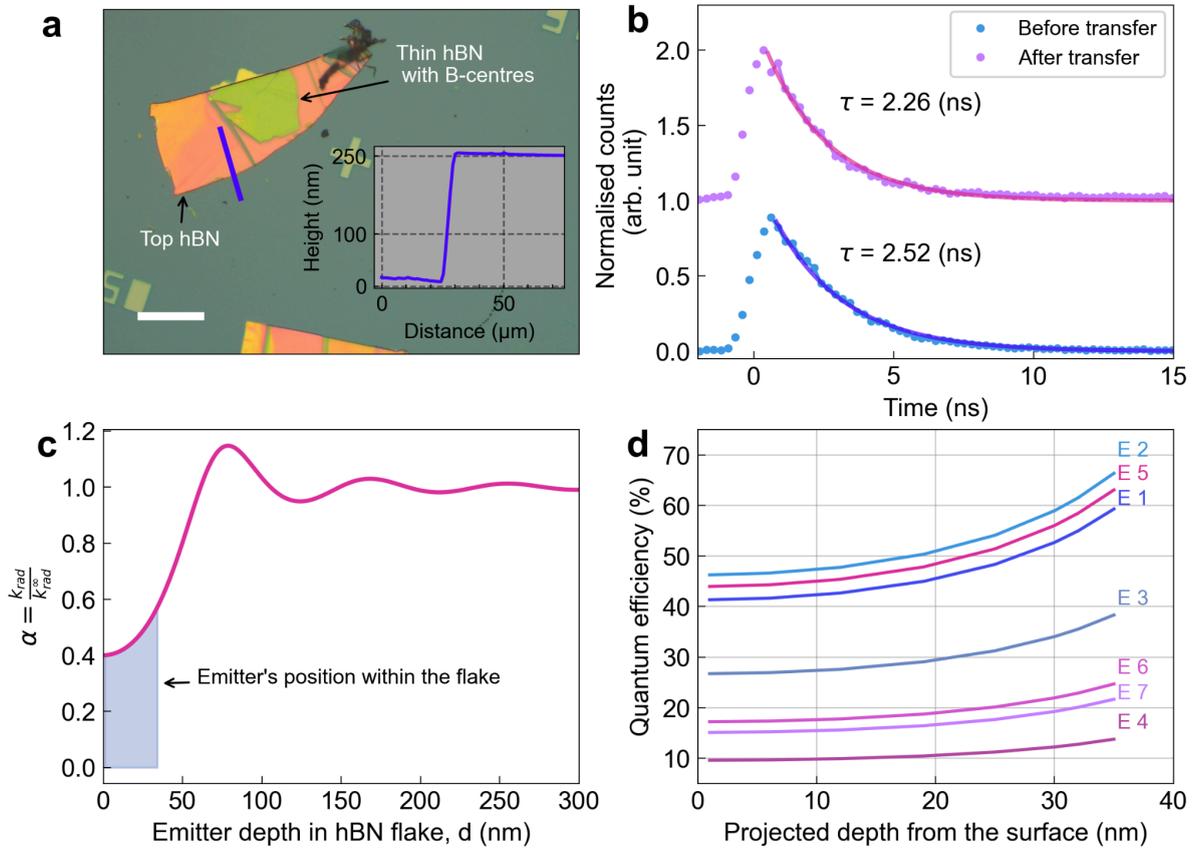

*Figure 2. (a) Optical image of the transferred thick flake of hBN on top of the 35 nm flake that hosts the B-centres. The inset is a height profile of the thick flake along the blue line. The scale bar is 50 µm. (b) Lifetime measurement from the same emitter performed before (light blue dots) and after (purple dots) transfer of the top hBN flake, performed using an excitation power of 50 µW. (c) Numerical simulation of the radiative decay rate ratio α versus the emitter depth in hBN flake. The shaded area shows the range of possible emitter depths within the 35 nm flake used in our experiments. (d) QE of the measured 7 emitters, calculated for depths ranging from 1 nm to 35 nm below the flake surface. Each curve corresponds to one emitter.*

Figure 2(a) shows an optical image of the 250 nm thick hBN flake, taken after it has been transferred on top of the 35 nm flake that contains the B-centres. The inset is an AFM height profile of the thick flake. Lifetime measurements of an emitter before and after the transfer of the additional flake on top are shown in Figure 2(b). The lifetime values are 2.52 (before transfer) and 2.26 ns (after transfer), respectively, for this particular B-centre. This measurement was then repeated for 7 emitters (data in the SI), yielding values ranging from 2.26 ns to 2.58 ns before transfer, and 1.73 ns to 2.26 ns after transfer, respectively. These values were used to deduce β for each emitter, while the α coefficient was obtained from numerical simulations.

The radiative decay rate is proportional to the radiative power of the dipole which is dependent on the depth d of the emitter below the surface and the refractive index. The rate of total radiative power at distance $d$ and $d= \infty$ was simulated using the classical dipole oscillator model, for an emitter near a hBN-air interface[23]. The simulated value is presented in Figure 2(c). The highest total power fraction is at a depth of ~ 50 nm, which validates the choice of a 35 nm flake in our experiment.

Since the penetration range of the employed electron beam was greater than 35 nm (i.e. the thickness of the original flake), the emitter generation process within the flake is stochastic. Consequently, the precise depth of each B-centre within the 35 nm flake is unknown. To circumvent this issue, we calculated the QE as a function of depth, as shown in Figure 2(d) for each of the 7 emitters (lifetime values of these emitters are shown in the SI). The mean QE value is $(28.6 \pm 14.0)\%$ at 1 nm, and it increases with depth due to modification of the LDOS close to the interface. When the emitter is positioned near the surface ($\frac{d}{\lambda} \ll 1$) and the refractive index of the second medium is lower than that of the first medium, the radiated power from the dipole is predominantly governed by the wide-angle interference of plane waves. However, as the separation distance d increases, the α value oscillates around 1. This phenomenon occurs because the evanescent waves don't reach the medium interface, thereby not contributing to the radiated power. The implication of this result is that the B-centres in principle should be bright enough to be observed even in a monolayer of a few layer hBN, having a projected QE of ~ 10% (E4 in our case).

While figure 2(d) shows the QE for an emitter near the interface, it is also important to estimate the QE after the transfer. Indeed, this would provide a closer estimation of the QE for B-centres in thick flakes of hBN, which typically show brighter and more stable emission, and therefore are more convenient for use in practical applications such as integration into nanostructures. Our data allow us to deduce this value : for an emitter initially at a depth d from the surface, the corresponding QE value calculated in Figure 2(d) allows to retrieve the radiative and non-radiative decay rates simply by applying equation (1) : $QE_1 = \frac{k_{rad,1}}{k_{rad,1}+k_{non-rad}}$ with $k_{rad,1} + k_{non-rad} = \frac{1}{\tau_1}$ where $QE_1$, $k_{rad,1}$ and $\tau_1$ denote respectively the QE, radiative decay rate and measured decay time before transfer. After the transfer, $k_{non-rad}$ is left unchanged and the measured decay time then allows to retrieve the new radiative decay rate : $k_{rad,2} + k_{non-rad} = \frac{1}{\tau_2}$ with the index 2 denoting the values after transfer. We can then apply equation (1) to obtain the quantum efficiency in the covered flake: $QE_2 = \frac{k_{rad,2}}{k_{rad,2}+k_{non-rad}}$ . The results are shown in Figure 3 for each of the 7 emitters, for an initial depth between 1 and 35nm. We obtain QE values above 42% for three out of seven emitters and a minimum QE value above 10%.

The values obtained in Figure 3 give an estimation of the QE for B-centres in thick flakes of hBN. A comparison in count rates between the characterised emitters in figure 2 and single B-centres in a reference thick flake at saturation power[26] could confirm these results. Indeed, saturation values of single B-centres in thick flakes are much higher than same B-centres in a thin flake, as shown in the SI.

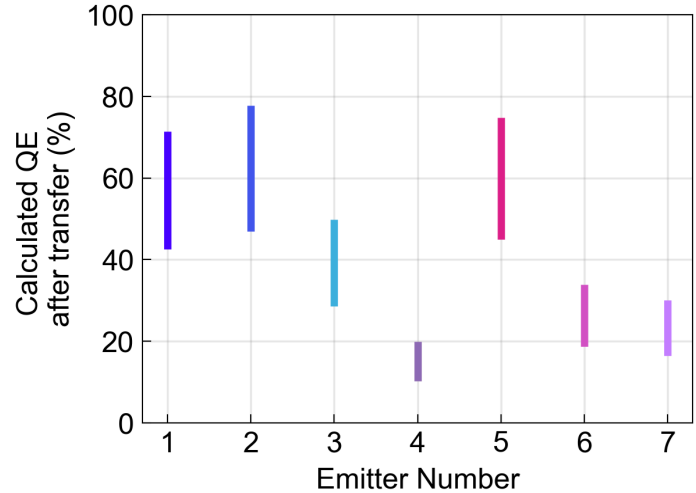

*Figure 3. Calculated QE for each emitter after transferring the top hBN flake. The interval for each emitter represents the uncertainty on its position within the flake.*

The reported values of QE for B-centres are in line with the reported values of QE for the visible emitters in hBN with ZPLs at 580 nm and 660 nm (62 ± 9% and 36 ± 8 % respectively)[19]. These values are promising for practical applications, and rank among the highest compared to other defects in solid-state materials such as group IV color centres in diamond or silicon carbide. Near-unity efficiency values are within reach by coupling the B-centre to a photonic cavity, increasing the radiative decay rate via the Purcell effect[27-29]. In particular, considering an initial QE of 50%, a Purcell factor of 9 would increase the QE above 90%. Overall, the combination of the fast radiative lifetime and high QE value of B-centres offers the perspective of single photon generation at GHz rates at room temperature. Such high brightness, combined with a reproducible ZPL emission wavelength and deterministic defect creation, make B-centres an attractive candidate as a bright single photon source for quantum photonic applications.

To conclude, we implemented an experimental method to measure the QE of single B-centres in hBN. We modified the radiative emission rate by transferring an additional thick hBN flake of 250 nm on top of the 35 nm hBN flake containing B-centres, leveraging the layered nature of the material. We measured lifetime changes in seven isolated B-centres before and after the transfer, and calculated their QE for a projected depth ranging from near the surface to 35 nm as well as after the transfer. Several emitters show QE values above 42% in the thin hBN flake, suggesting potential for near-unity QE under Purcell enhancement. Our work characterises an important figure of merit for the B-centres in hBN, and highlights their promise for quantum photonics applications.

During the preparation of our manuscript, we became aware of a related study, reporting near-unity QE for B-centres in hBN with thickness from 70 to 220 nm[30], which are in agreement with our results.

## Acknowledgements

This research is supported by the Australian Research Council (CE200100010, FT220100053) and the Office of Naval Research Global (N62909-22-1-2028). The authors thank the ANFF node of UTS for access to facilities. We acknowledge Takashi Taniguchi (the National Institute for Materials Science (NIMS)) for providing hBN crystal.

**Materials and Methods**

Sample preparation and emitter activation

Silicon substrates were cleaned using the sonication method in acetone and isopropanol for 10 minutes, respectively. The substrates were dried under nitrogen flow and placed in an air plasma chamber for 10 minutes. Carbon-doped hBN bulk crystals provided by the National Institute of Materials Science (NIMS) were exfoliated using scotch tape. Flake thickness was confirmed by AFM measurement using a Park XE7 atomic force microscope. The samples were annealed in air for 3 hours at 500 °C and subsequently cleaned using in-situ air plasma cleaning in the SEM chamber for 20 minutes.

The SPEs in hBN were activated using electron beam irradiation using a ThermoFisher G4 dual-beam microscope, Arrays of B-centres were patterned on the 35 nm hBN flake using electron beam irradiation at 3 keV and 1.6 nA with a dose range from $1.5 \times 10^9$ to $6.0 \times 10^9$ electrons. Similarly, arrays of B-centres were activated in a reference hBN flake with 250 nm thickness for the saturation measurement in Figure 3. After cleaning, an additional h-BN flake (fig 2(a)) was transferred on top of the thin flake to effectively use it as a dielectric medium to modulate the radiative decay rates. The flake transfer process involved using polypropylene carbonate (PPC) polymer, which was subsequently placed in a beaker with acetone for 10 minutes to remove any polymer residue.

The sample was exposed to UV Ozone for 30 minutes prior to the optical measurements to reduce the background signal in the PL that arose from hydrocarbon contaminants around the irradiated region.

Optical characterisation

Optical characterization included lifetime measurement, PL spectra, polarisation, and second-order correlation measurements. The schematics for the measurement is shown in Fig. 1(c), a 405 nm pulsed laser focused with an objective lens (100x/0.90 NA, Nikon TU Plan Fluor) onto the sample The SPE fluorescence is collected through a confocal setup into a Hanbury Brown and Twiss setup consisting of a polarising beam splitter (PBS) and two avalanche photodiodes (APD) (Excelitas). For polarisation measurement, a λ/4 wave plate was placed on the excitation path, and λ/2 and polarisation beam splitter were placed on the collection path. Additionally, a 405 nm continuous wave (CW)(PiL040X, A.L.S. GmbH) laser was employed for the saturation measurement.